\documentclass[preprint]{revtex4-1}
\usepackage{mathtools}
\usepackage{bm}
\usepackage{amsmath,amssymb}
\usepackage{color}
\usepackage{braket}
\usepackage{graphicx}
\usepackage{epstopdf}
\usepackage{float}
\usepackage{nccmath}
\usepackage[normalem]{ulem} %\sout
\usepackage{setspace}
\newcommand\T[1]{\ignorespaces} % comment this part

\begin{document}
\title{Winding around Non-Hermitian Singularities: General Theory and Topological Features}

%% Notice placement of commas and superscripts and use of &
%% in the author list

\author{Q. Zhong$^1$}
\author{M. Khajavikhan$^2$}
\author{D.N. Christodoulides$^2$}
\author{R. El-Ganainy$^{1,3}$ }
\email[]{ganainy@mtu.edu}

\affiliation{$^1$ Department of Physics, Michigan Technological University, Houghton, Michigan, 49931, USA}
\affiliation{$^2$ College of Optics $\&$ Photonics-CREOL, University of Central Florida, Orlando, Florida, 32816, USA}
\affiliation{$^3$ Center for Quantum Phenomena, Michigan Technological University, Houghton, Michigan, 49931, USA}

\begin{abstract}
Non-Hermitian singularities are ubiquitous in non-conservative open systems. These singularities are often points of measure zero in the eigenspectrum of the system which make them difficult to access without careful engineering. Despite that, they can remotely induce observable effects when some of the system's parameters are varied along closed trajectories in the parameter space. To date, a general formalism for describing this process beyond simple cases is still lacking. Here, we bridge this gap and develop a general approach for treating this problem by utilizing the power of permutation operators and representation theory. This in turn allows us to reveal the following surprising result which contradicts the common belief in the field: loops that enclose the same singularities starting from the same initial point and traveling in the same direction, do not necessarily share the same end outcome. Interestingly, we find that this equivalence can be formally established only by invoking the topological notion of homotopy. Our findings are general with far reaching implications in various fields ranging from photonics and atomic physics to microwaves and acoustics.	
\end{abstract}

\maketitle
\subsection*{Introduction}
Non-Hermitian singularities arise in multivalued complex functions \cite{Needham-VCA,Ablowiz-CV} as points where the Taylor series expansion fails. In the context of non-Hermitian Hamiltonians, these points, commonly referred to as exceptional points (EPs) feature special degeneracies where two or more eigenvalues along with their associated eigenfunctions become identical \cite{Heiss2004JPA,Muller2008JPA}. An EP of order $N$ (EPN) is formed by $N$ coalescing eigenstates. Recently, the exotic features of EPs have been subject of intense studies \cite{El-Ganainy2007OL,El-Ganainy2008PRL-O,El-Ganainy2008PRL-B,El-Ganainy2010NP}  with various potential applications in laser science \cite{Hodaei2014S,Feng2014S, El-Ganainy2015PRA,Teimourpour2016SR} , optical sensing \cite{Wiersig2014PRL, Hodaei2017N,Chen2017N},  photon transport engineering \cite{Lin2011PRL,Zhu2013OL} and nonlinear optics \cite{El-Ganainy2015OL,Zhong2016NJP}  just to mention few examples. For recent reviews, see Refs \cite{El-Ganainy2018NP,Feng2017NP}. 

Very often, EPs are points of measure zero in the eigenspectra of non-Hermitian Hamiltonians which makes them very difficult to access, even with careful engineering. Yet, their effect can be still felt globally. Particularly, an intriguing aspect of non-Hermitian systems is the eigenstate exchange along loops that trace closed trajectories around EPs. In this regard, stroboscopic encircling of EP2 has been studied theoretically \cite{Heiss1999EPD,Cartarius2007PRL} and demonstrated experimentally in various platforms such as microwave resonators \cite{Dembowski2001PRL,Dietz2011PRL} and exciton-polariton setups \cite{Gao2015N}. Complementary to these efforts, the dynamic encircling of EPs was shown to violate the standard adiabatic approximation \cite{Raam2011JPA,Berry2011JPA,Berry2011JO,Hassan2017PRL}. These predictions were recently confirmed experimentally by using microwave waveguides platforms \cite{Doppler2016N} and optomechanical systems\cite{Xu2016N}. 

Notably, the aforementioned studies focused only systems having only one EP of order two. Richer scenarios involving multiple and/or higher order EPs have been largely neglected, with rare exceptions that treated special systems (admitting simple analytical solutions) on a case by case basis \cite{Ryu2012PRA, Demange2012JPA}. This gap in the literature is probably due to the complexity of the general problem and its perceived experimental irrelevance. However, recent progress in experimental activities that explore the physics of non-Hermitian systems are quickly changing the research landscape, and controlled experiments that probe more complicated structures with multiple EPs will be soon within reach. These developments beg for a general approach that can provide a deeper theoretical insight into these complex systems.

In this work, we bridge this gap by introducing a general formalism for treating the eigenstate exchange along arbitrary loops enclosing multiple EPs. More specifically, our approach utilizes the power of group theory together with group representations to decompose the final action of any loop into more elementary exchange processes across the relevant branch cuts (BCs). This formalism simplifies the analysis significantly, which in turn allows us to gain an insight into the problem at hand and unravel a number of intriguing results: (1) Trajectories that encircle the same EPs starting from the same initial point and having the same direction do not necessary lead to an identical exchange between the eigenstates; (2) Establishing such equivalence between the loops (i.e. same eigenstate exchange) is guaranteed only by invoking the topological notion of homotopy. As a bonus, our approach can also paint a qualitative picture of the dynamical properties of the system. 

\subsection*{General Formalism for Encircling Multiple Exceptional Points}
\begin{figure}
	\includegraphics[width=\linewidth*3/4]{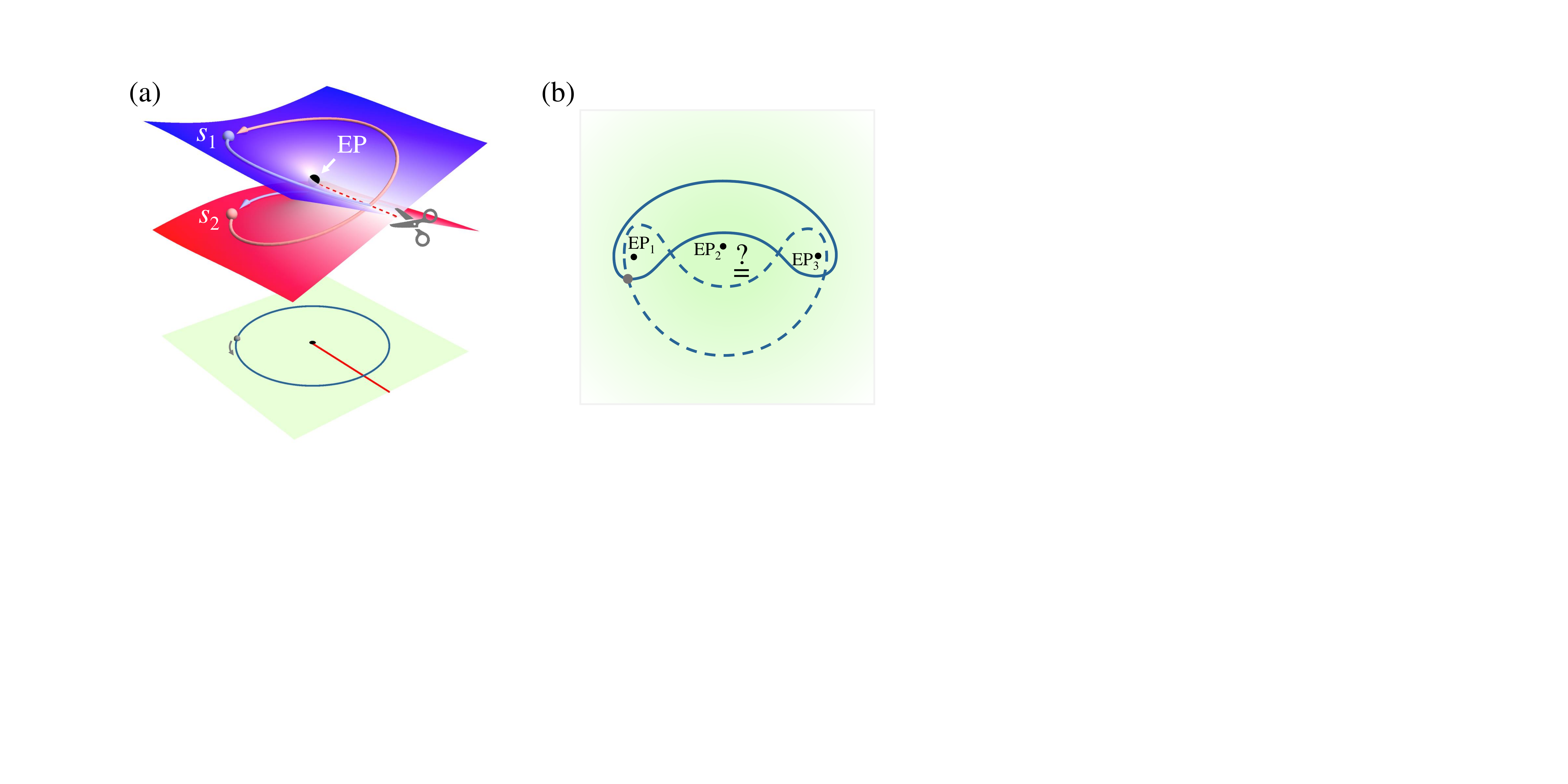}
	\caption{\textbf{Different ways of encircling multiple EPs.} (a) Illustration of Riemann surface associated with the square root function associated with an archetypal $2 \times 2$ non-Hermitian Hamiltonian. A loop that encircles the exceptional point (also known as the branch point) starting from the state $s_1$ will map it onto $s_2$ and vice versa. In the complex plane (lower panel), this is represented by adding a BC. (b) A scenario that exhibit three EPs. In this case, loops can encircle EPs in different ways as illustrated by the two loops (solid/dashed lines) that enclose $\text{EP}_{1,3}$ starting from the same point (gray dot).    
	}
	\label{FigEPs}
\end{figure}

Before we start our analysis, we first describe the simple case of EP2. These are special points associated with the multivalued square root function in the complex plane. The Riemann surface of this function is shown in the top panel of Fig. \ref{FigEPs}(a). Clearly, as two parameters are varied in the complex plane to trace a closed loop, the initial point on the surface ends up on a different sheet. This process can be also viewed by considering the projection on the complex plane after adding a BC (lower panel). As we mentioned before, this simple scenario has been studied in the literature in both the stroboscopic and dynamical cases. Consider however what happens in more complex situations where there are more than one EP. For instance, Fig. \ref{FigEPs}(b) depicts a case with three EPs.  One can immediately see that this scenario exhibits an additional complexity that is absent from the previous case. Namely, there are now different ways for encircling the same EPs (as shown by the solid and dashed loops in the figure). This in turn raises the question as whether these loops lead to the same results or not. These are the type of questions that we would like to address in this work. As we will see, in resolving these questions, our analysis also reveal several peculiar scenarios.

\begin{figure}
	\includegraphics[width=\linewidth*3/4]{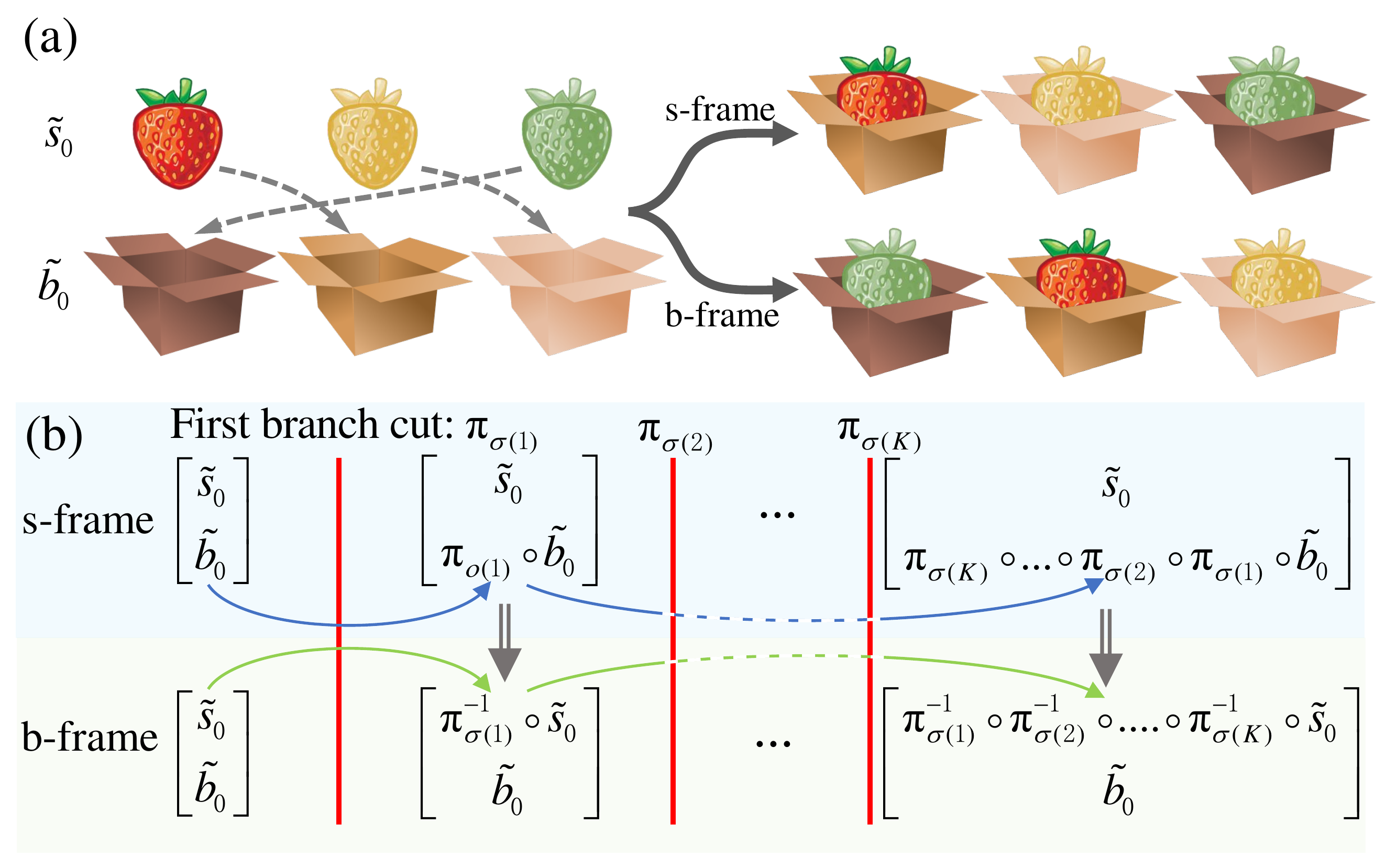}
	\caption{\textbf{Different permutation frames.} (a) A simple illustration of the two different frames used for representing the same configuration. (b) The mathematical formulation of the concept in (a) in terms of permutation mappings as discussed in details in the text.}
	\label{FigPermutation} 
\end{figure}

\noindent \textit{Permutation operators and the exchange of eigenstates---}
Consider an $n$-dimensional non-Hermitian discrete Hamilton. The Riemann surface associated with the real (or imaginary) part of its eigenvalues will consist of $n$ sheets corresponding to different solution branches. We will label these $n$ branches as $b_1$, $b_2$, ..., $b_n$. In the complex plane, these branches are separated by BCs. Thus, an initial point on any trajectory in the complex plane will correspond to $n$ initial eigenstates, which we will label as $s_1$, $s_2$, ..., $s_n$. The eigenvalue for each state $s_i$ will be denoted by $\lambda_i$.  As the encircling parameters are varied, the eigenstates will move along the trajectory, crossing from one branch to another across the BCs. The crucial point here is that, we will always fix the initial subscript of the state as it changes. We now describe the initial configuration on the trajectory by the mapping:
\begin{equation}
\label{Eq:C0}
\mathcal{C}_0=
\begin{bmatrix}
\tilde{s}_0 \\
\tilde{b}_0 
\end{bmatrix},
\end{equation}
where $\tilde{s}_0=(s_1 , s_2,...,s_n)$ and  $\tilde{b}_0=(b_1 , b_2 , ..., b_n)$ are two ordered sets. In our notation, $\mathcal{C}_0$ maps (or associates) every element of $\tilde{s}_0$ to the corresponding element in $\tilde{b}_0$. Note that we can change the orders of the elements in both $\tilde{s}_0$ and $\tilde{b}_0$ identically without changing $\mathcal{C}_0$. In other words, we have several different ways for the same configuration. As the loop crosses BCs, the exchange between the eigenstates will result in new configurations which, again, can be described in different ways. Two particular choices are interesting here. In the first one, we always fix $\tilde{s}_0$ and allow the elements of $\tilde{b}_0$ to shuffle, effectively creating a new $\tilde{b}$. In the second, we just do the converse. We will call these two equivalent notations the s- and b-frames, respectively. This is explained by the cartoon picture in Fig. \ref{FigPermutation}(a).

The first step in our analysis is to choose a scheme for sorting the eigenstates and locating the BCs accordingly. We will discuss the details of the sorting later but for now we assume that we have a certain number of BCs and we label each one with a unique integer value (positive for a crossing in certain direction and negative for reverse crossing). Next we determine how the eigenstates are redistributed across an infinitesimal trajectory across each BC (see discussion later on sorting schemes). For every loop, we then create an ordered list $\sigma$ that contains the number of the crossed BCs in the order they are crossed by the loop. In other words, the element $\sigma(j)$ is the number of the $j$-th crossed BC. Clearly the set $\sigma$ will be in general different from loop to another and even can be different for the same loop depending on the initial point or the encircling direction. Then the final configuration in both the s- and b-frames is given by:
\begin{equation}\label{Eq:Cm}
\begin{aligned}
\mathcal{C}^\text{s}_\sigma&=
\begin{bmatrix}
\tilde{s}_0 \\
\tilde{b}_\sigma 
\end{bmatrix}
\equiv
\begin{bmatrix}
\tilde{s}_0 \\
\mathcal{P} [\prod \pi_{\sigma(j)}] \circ \tilde{b}_0 
\end{bmatrix},\\
\mathcal{C}^\text{b}_\sigma&=
\begin{bmatrix}
\tilde{s}_\sigma \\
\tilde{b}_0 
\end{bmatrix}
\equiv
\begin{bmatrix}
\{\mathcal{P} [\prod \pi_{\sigma(j)}]\}^{-1} \circ \tilde{s}_0 \\
\tilde{b}_0 
\end{bmatrix},
\end{aligned}
\end{equation}	
where $\mathcal{P}$ denotes the ordering operator which arranges the multiplication of the permutation operators $\pi_{\sigma(j)}$ from right to left according to the order of crossing the BCs; and the product runs across the index $j$. For example, if $\sigma=(3,1,2)$, then the $\mathcal{P} [\prod \pi_{\sigma(j)}]=\pi_{\sigma(3)} \circ \pi_{\sigma(2)} \circ \pi_{\sigma(1)}=\pi_2 \circ \pi_1 \circ \pi_3$. The permutation operator $\pi_{k}$ associated with BC $k$ is the standard permutation mapping that, which when applied to a set will shuffles the order of its elements \cite{Hassani-MP}. Here it is used to describe how the eigenstates are redistributed when a trajectory crosses a BC. For instance, if the permutation exchange the order of the first two elements of $\tilde{b}_0$ across a BC $k$, then $\pi_k(b_{1,2})=b_{2,1}$, and $\pi_k(b_i)=b_i$ for $i>2$. Figure \ref{FigPermutation}(b) illustrates the relation between the s- and b-frame calculations as expressed by Eqs. (\ref{Eq:Cm}).

\noindent \textit{From permutations to matrices---}
The above discussion can be directly mapped into linear algebra by using representation theory. To do so, we define the vectors $\vec{s}_0=(s_1,s_2, ..., s_n)^T$  and $\vec{b}_0=(b_1,b_2, ..., b_n)^T$. In the s-frame, we will fix $\vec{s}_0$ and allow $\vec{b}$ to vary in order to represent the change in configuration. In the b-frame, we just do the opposite. For instance, if after crossing a BC, eigenstate 1 moves to branch $n$, eigenstate 2 moves to branch 1 and eigenstate $n$ moves to branch 2, this will be expressed as $\vec{b}_1=(b_n,b_1, ..., b_2)^T$ in the s-frame; and $\vec{s}_1=(s_2,s_n,...s_1)^T$ in the b-frame.  After a loop completes its full cycle,  the final vector is then compared with the initial one to determine the exchange relations between the eigenstates. For instance, if the above vector was the final result, the exchange relations will be: $\{s_1,s_2,...,s_n\} \rightarrow \{s_n, s_1,...,s_2\}$, which means that after the evolution $s_1$ became $s_n$, $s_2$ became $s_1$ and $s_n$ became $s_2$. 

We can now express the action of the permutation operators $\pi_k$ by the matrices $\textbf{\text{P}}_{\pi_k}$ whose elements are obtained according to the rule $\mathbf{P}_{\pi_k}(m,l) = 1$ if $b_l = \pi_k(b_m)$, and 0 otherwise \cite{Brualdi-CMC}. In the s- $\&$ b-frames, the redistribution of the eigenstates across the branches in Eq. (\ref{Eq:Cm}) can be then described by:
\begin{equation} \label{Eq: Matrix product}
\begin{aligned}
\vec{b}_\sigma &=\{\mathcal{P}[\prod \mathbf{M}_{{\sigma(j)}}]\}^{-1}\vec{b}_0, \\
\vec{s}_\sigma &=\mathcal{P}[\prod \mathbf{M}_{{\sigma(j)}}]\vec{s}_0,
\end{aligned}
\end{equation} 
where $\mathbf{M}_k=\mathbf{P}_{\pi_k}^{-1}$. In arriving at the above equation, we have used standard results from group theory: $\mathbf{P}_{\pi_2 \circ \pi_1}=\mathbf{P}_{\pi_1} \mathbf{P}_{\pi_2}$ and $\mathbf{P}_{\pi^{-1}}=\mathbf{P}_{\pi}^{-1}$. 

In the rest of this manuscript, we adopted the b-frame with matrices $\mathbf{M}$. This approach offers a clear advantage: the order of the matrices acting on the state vectors $\vec{s}$ is consistent with the order of crossing the BCs. As we will see shortly, this will allow us to develop the topological features of the equivalent loops in a straightforward manner. Finally, we note that if crossing a BC from one direction to another is associated with a matrix $\mathbf{M}$, the reverse crossing will be described by $\mathbf{M}^{-1}$. In some cases (such as with EP2), we can have $\mathbf{M}^{-1}=\mathbf{M}$ but this is not the general case.

\noindent \textit{Sorting of the eigenstates---} The discussion so far focused on developing the general formalism by assuming that the eigenstates of the system are somehow classified according to a certain criterion. This is equivalent to say that we divide the associated Riemann surface into different sheets, each harboring a solution branch. Of course, one can pick any such criterion to classify the solutions. In previous studies that involved one EP of order two or three, the eigenstates were classified based on the analytical solution of the associated characteristic polynomial. This however has two drawbacks: (1) It generates relatively complex branches on the Riemann sheet; (2) It cannot be applied for discrete Hamiltonians having dimensions larger than four since analytical solutions do not exist for polynomials of order five or larger. Thus our analysis above is useful only if one can find a sorting scheme that circumvents the above problems. Interestingly, such a sorting scheme is easy to find. Particularly, we can sort the eigenstates based on the ascending (or descending) order of the real or imaginary parts of their eigenvalues. This scheme can be easily applied to any system of arbitrarily high dimensions. Moreover, it lends itself to straightforward numerical implementations. To compute the a permutation operator $\pi_k$ and its associated matrix $\mathbf{M}_k$ across a BC $k$, one choses an infinitesimal trajectory that crosses the BC and calculate how the eigenvalues evolve along this trajectory, comparing their order before and after crossing the BC. That will immediately provide information about the permutations. We illustrate this using concrete example in the Methods.  

\subsection*{Equivalent Loops and Homotopy}
In this section, we employ the predictive power of our formalism to address the following question: are there any global features that characterize the equivalence between different loops regardless of their geometric details? In answering this question, we will first focus on the stroboscopic case and later discuss the implication for the dynamical behavior.  

Here, two loops are called equivalent if they lead to identical static eigenstates exchange. It is generally believed that two similar loops starting at the same point and encircling the same EPs in the same direction are equivalent. Surprisingly, we will show below that this common belief is wrong. 

In general two loops will be equivalent if they have the same matrix product in Eq. (\ref{Eq: Matrix product}). This can occur for two unrelated loops which we will call accidental equivalence. However, We are particularly interested in establishing the conditions that guarantee this equivalence. To do so, we invoke the notion of homotopy between loops. In topology, two simple paths, having the same fixed endpoints in a space $S$, are called homotopic if they can be continuously deformed into each other \cite{Hatcher-AT}. Here the word ``simple" means injective, that is, each path does not intersect itself. If the two endpoints of a path are identical, this path is a loop with the identical endpoint as a basepoint. The space $S$ here will be a two dimensional punctured parameter space (for example, the space spanned by Re$[\kappa]$--Im$[\kappa]$ in the examples discussed in the Methods) after removing all the EPs. Based on these definitions, we can now state the main results of this section: \textbf{(a) Homotopy is a sufficient condition for equivalence between loops; (b) Loops that are connected by free homotopy (continuous deformation between loops without any fixed points) can be equivalent for some starting points and inequivalent for others}.

\begin{figure}
	\includegraphics[width=\linewidth]{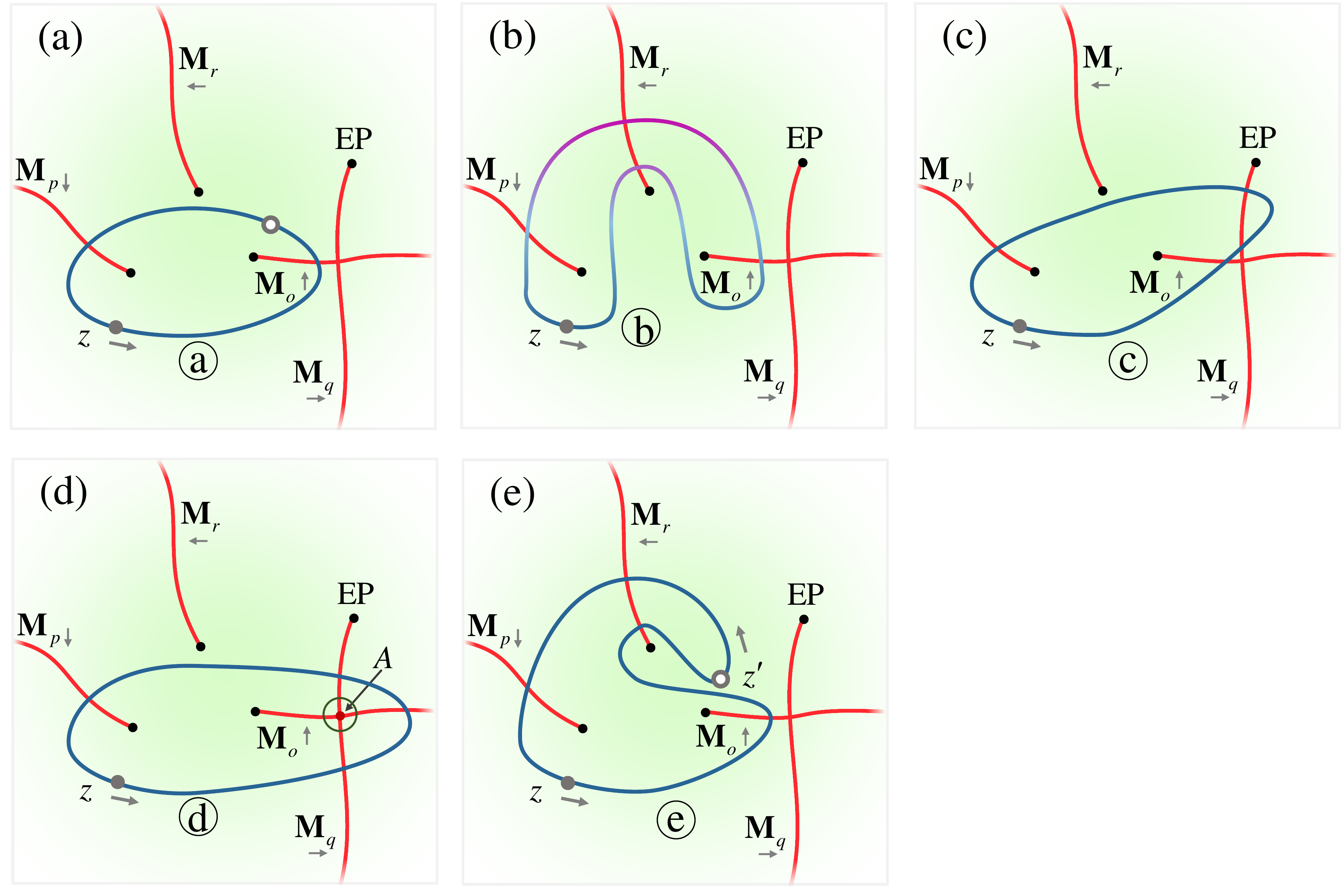}
	\caption{\textbf{Homotopy between loops.} Illustration of equivalence between homotopic loops in the parameter space of a generic Hamiltonian. (a) Loop \textcircled{a} encloses two EPs associated with matrices $\mathbf{M}_o$ and $\mathbf{M}_p$. (b) Loop \textcircled{b} encloses the same two EPs yet it cannot be deformed into \textcircled{a} without crossing EP associated with $\mathbf{M}_r$. Consequently it has different matrix product (assuming not accidental equivalence). On the other hand, loops \textcircled{c} and \textcircled{d} in (c) and (d) can be deformed into \textcircled{a} without crossing any EP. As a result, they are equivalent (have the same matrix product) as shown in the text. (d) Presents a peculiar case of free homotopy. Loop \textcircled{e} is homotopic with \textcircled{a} for the starting point $z$ but not for $z'$. As a result, the two loops are equivalent for the former point but not for the latter. The discussion here is very generic and can be extended easily to any other configuration of EPs and BCs.}
	\label{FigHomotopy} 
\end{figure} 

In order to validate this statement, we consider a generic Hamiltonian having a number of EPs and, without any loss of generality, we focus only on a subset of the spectrum as shown in Fig. \ref{FigHomotopy}. The axes on the figures represent any two parameters of the Hamiltonian. We define the space $S$ to be the two dimensional parameter space excluding the EPs. Figure \ref{FigHomotopy}(a) depicts a loop \textcircled{a} that encircles two EPs starting from point $z$ in the counterclockwise(CCW) direction. Consequently, its final permutation matrix is given by $\mathbf{M}_p\mathbf{M}_o$. Consider now what happens when loop \textcircled{a} is deformed continuously to a new loop. Here different scenarios can arise: (1) The deformation can take place only by crossing additional EP any number of times. This case is shown in Fig. \ref{FigHomotopy}(b), where it is clear that the new matrix product of loop \textcircled{b} ($\mathbf{M}_p\mathbf{M}_r\mathbf{M}_o\mathbf{M}_r^{-1}$) is in general different than the initial one. In this case, the two loops are not equivalent (unless accidental equivalence takes place). (2) The deformation can occur without changing the number or order of the crossed BCs, in which case the loops are equivalent. (3) The deformation can change the number of the crossed  BCs in pairs traversed consecutively back and forth as shown in  Fig. \ref{FigHomotopy}(c). Here the two loops \textcircled{a} and \textcircled{c} are also equivalent because the matrix product is still the same: $\mathbf{M}_p\mathbf{M}_q^{-1}\mathbf{M}_q\mathbf{M}_o=\mathbf{M}_p\mathbf{M}_o$. (4) The deformation can occur without crossing any EP but it changes the number of the crossed BCs in pairs traversed back and forth but not consecutively as shown in Fig. \ref{FigHomotopy}(d). In this case, the final matrix product is given by $\mathbf{M}_p\mathbf{M}_q^{-1}\mathbf{M}_o\mathbf{M}_q$. It is not immediately clear if this product is equivalent to $\mathbf{M}_p\mathbf{M}_o$. However, since the intersection point of the BCs (point $A$) is not an EP, then by definition, encircling point $A$ with a loop that does not enclose any EP must give the identity operator. In terms of matrices,  this translates into $\mathbf{M}_o \mathbf{M}_q \mathbf{M}_o^{-1} \mathbf{M}_q^{-1} =\mathbf{I}$, or $[\mathbf{M}_o , \mathbf{M}_q]=0$. Consequently, $\mathbf{M}_p\mathbf{M}_q^{-1}\mathbf{M}_o\mathbf{M}_q=\mathbf{M}_p\mathbf{M}_o\mathbf{M}_q^{-1}\mathbf{M}_q=\mathbf{M}_p\mathbf{M}_o$, i.e. loops \textcircled{d} and \textcircled{a} are equivalent. (5) Finally we can also have a loop similar to \textcircled{e} as shown in Fig. \ref{FigHomotopy}(e). This probably the most intriguing situation. For a starting point at $\kappa_0$, both loops \textcircled{a} and \textcircled{e} have the same matrix product $\mathbf{M}_p\mathbf{M}_o$ which is consistent with the fact that they can be deformed into one another without crossing any EP. On the other hand, for a different starting point such as $z'$, the matrix product of loop \textcircled{e} is given by $\mathbf{M}_r^{-1} \mathbf{M}_o \mathbf{M}_p \mathbf{M}_r$, i.e. different than that of loop \textcircled{a}, which is given by $\mathbf{M}_o \mathbf{M}_p$. Note that for this starting point, the two loops cannot be deformed into each other without crossing any EP. In topology, continuous deformation that do not involve fixed points are called free homotopy. This completes our argument.

The above discussion focused only on the stroboscopic case. However, as we will show in the explicit example presented in Methods, homotopy is also relevant to the dynamical encircling of EPs. Particularly, our numerical calculations show that homotopic loops tend to have the same outcome, despite the failure of the adiabatic perturbation theory. Intuitively, this interesting result can be roughly understood by noting that homotopic loops explore very similar landscape in the complex domain. However a deeper understanding of this behavior requires further investigation.

\subsection*{Conclusion}
In conclusion, we have introduced a general formalism based on permutation groups and representation theory for describing the stroboscopic encircling of multiple EPs. By using this tool, we  uncovered the following counterintuitive results: trajectories that enclose the same EPs starting from the same initial parameters and  traveling in the same direction,  do not necessarily result in identical exchange between the states. Instead, we have shown that this equivalence can be established only between homotopic loops. Finally we have also discussed the implication of these results for the dynamic encircling of EPs. Our work may find applications in various fields including the recent interesting work on the relationship between exceptional points and topological edge states \cite {leykam2017PRL, Hu2017PRB}.

\subsection*{Method}
\noindent\textit{Illustrative Examples ---}
We now discuss a concrete numerical example to demonstrate the application of our formalism and confirm the various predictions presented in the main text.

\noindent \textit{Model---} Consider the following Hamiltonian:
\begin{equation}\label{H}
\renewcommand\arraystretch{0.75}
H=\begin{bmatrix}
i \gamma & J & 0 & 0 \\
J & 0 & \kappa & 0 \\
0 & \kappa & 0 & J \\
0 & 0 & J & -i \gamma
\end{bmatrix},
\end{equation}
where $i$ is the imaginary unit, $\kappa$ $\&$ $J$ are coupling coefficients and $\gamma$ is the non-Hermitian parameter. In what follows, the four eigenvalues of $H$ will be investigated as a function of the complex $\kappa$ by fixing $J=\gamma=1$ (in certain physical platforms such as optics, it might be practically easier to fix all the parameters and change $\gamma$, but that will not affect the main conclusions of this work). 

\begin{figure} 
	\includegraphics[width=\linewidth]{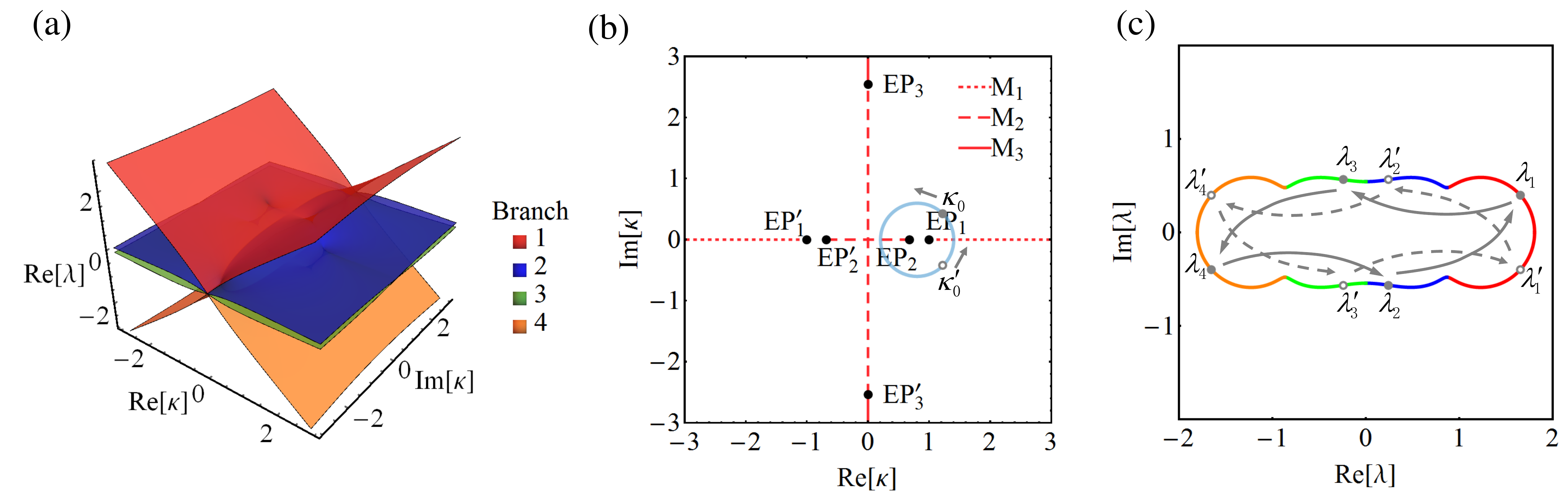}
	\caption{\textbf{Numerical illustration of our approach.} (a) The branches of Riemann surface of the real part of eigenvalues of $H$ in Eq. (\ref{H}) are distinguished by different colors according to the magnitude of Re[$\lambda$]. The EPs and their corresponding BCs (red lines) are illustrated in (b). Each BC is related with a permutation matrix $\mathbf{M}_k$ in Eq. (\ref{Matrix}). One closed loop (blue line) encircles EP$_1$ and EP$_2$ CCW, starting from the gray points (solid or hollow) on the loop. Loops intersecting with BCs would lead to eigenvalues moving from one branch to another, and result in the swap of eigenstates finally. (c) The stroboscopic evolution of complex eigenvalues are plotted as a parametric function of $\kappa$ when it moves along the loop CCW. The eigenvalues at the starting point are labeled as gray points on their trajectory. The colors in the eigenvalue trajectory represent which branch the eigenvalues are located at instantaneously. The joints of two colors are where the $\kappa$ crosses the BCs.  The gray points  (solid or hollow)  and arrows illustrate the evolution of eigenvalues for starting from $\kappa_0$ or $\kappa'_0$, and therefore the evolution of eigenstates is $\{s_1,s_2,s_3,s_4\} \rightarrow \{s_3,s_1,s_4,s_2\}$ and  $\rightarrow\{s_2,s_4,s_1,s_3\}$, respectively.} 
	\label{FigExample1} 
\end{figure}

Under these conditions, $H$ has three pairs of EPs at $\kappa=\pm1$, $\pm\sqrt{2\sqrt{3}-3}$, $\pm i \sqrt{2\sqrt{3}+3}$, which we will denote by EP$_1$, EP$_1'$, EP$_2$, EP$_2'$, EP$_3$, EP$_3'$, respectively. In each group, EP$_{1,2,3}'$ has same properties as EP$_{1,2,3}$. The Riemann surface and the distribution of the EPs in the complex $\kappa$ plane are shown in Fig. \ref{FigExample1}(a) and (b), respectively.

As discussed in the main text, the first step in our approach is to identify a simple sorting method. Here we chose to sort the eigenvalues according to the magnitude of their real parts as shown in Fig. \ref{FigExample1}(a) where every branch is distinguished by a distinct color. From this figure, we can also identify the features of the EPs as follows: EP$_1$ $\&$ EP$_1'$ are of second order and connect branches 2 and 3; EP$_2$ $\&$ EP$_2'$ are of second order and connect branches 1 and 2 on one hand, and branches 3 and 4 on the other; and finally EP$_3$ $\&$ EP$_3'$ are of second order and connect branches 1 and 3 as well as branches 2 and 4 (In fact all the four surfaces of Re$[\lambda]$ are connected at EP$_3$ $\&$ EP$_3'$ and one has to look at the Im$[\lambda]$ surface, which is not shown here, to infer the connectivity). Equivalently, the surface connectivity across the EPs can be characterized by using a two dimensional plane spanned by the real and imaginary parts of $\kappa$ along with the lines that separate the different solution branches (BCs) and the information on the transition between the different branches across each line. The latter can be expressed in terms permutation matrices. Our sorting scheme of the eigenvalues of $H$ results in six BCs as shown in Fig. \ref{FigExample1}(b), but one can identify only three different permutation matrices: 
\begin{equation}\label{Matrix}
\renewcommand\arraystretch{0.5}
\begin{split}
\mathbf{M}_1&=\begin{bmatrix}
1 & 0 & 0 & 0\\
0 & 0 & 1 & 0\\
0 & 1 & 0 & 0\\
0 & 0 & 0 & 1 
\end{bmatrix}, 
\mathbf{M}_2=\begin{bmatrix}
0 & 1 & 0 & 0\\
1 & 0 & 0 & 0\\
0 & 0 & 0 & 1\\
0 & 0 & 1 & 0 
\end{bmatrix}, 
\mathbf{M}_3=\begin{bmatrix}
0 & 0 & 0 & 1\\
0 & 0 & 1 & 0\\
0 & 1 & 0 & 0\\
1 & 0 & 0 & 0 
\end{bmatrix}. 
\end{split}
\end{equation}
The correspondence between these matrices and the BCs is depicted in Fig. \ref{FigExample1}(b).   It is not difficult to see that the above matrices have the following properties: (1)  $\mathbf{M}_1^2=\mathbf{M}_2^2=\mathbf{M}_3^2=\mathbf{I}$; (2) $[\mathbf{M}_1, \mathbf{M}_3]=[\mathbf{M}_2, \mathbf{M}_3]=0$.

\noindent \textit{Stroboscopic encircling of EPs---} We now focus on the loop encircling both EP$_1$ and EP$_2$,  as shown in Fig. \ref{FigExample1}(b). Clearly, the final exchange relation is determined by the product of $\mathbf{M}_1$ and $\mathbf{M}_2$. Since $[\mathbf{M}_1,\mathbf{M}_2] \neq 0$, one has to be more specific about the starting point and direction. For sake of illustration, let us choose counterclockwise direction, and  $\kappa_0$ or $\kappa'_0$ as the starting point. In the first case, the loop intersects the BC associated with $\mathbf{M}_2$ first before it crosses that of $\mathbf{M}_1$. As such, we have $\mathbf{M}_1 \mathbf{M}_2 (s_1, s_2, s_3, s_4)^T=(s_2, s_4, s_1, s_3)^T$, which in turn implies the exchange $\{s_1, s_2, s_3, s_4\} \rightarrow \{s_3, s_1, s_4, s_2\}$. Similarly, the starting point $\kappa_0^{\prime}$ will give  $\mathbf{M}_2 \mathbf{M}_1 (s_1, s_2, s_3, s_4)^T=(s_3, s_1, s_4, s_2)^T$ which leads to $\{s_1, s_2, s_3, s_4\} \rightarrow \{s_2, s_4, s_1, s_3\}$. These exchange relations are also evident from the eigenvalues trajectories in Fig. \ref{FigExample1}(c).  Another important consequence for the absence of commutation between $\mathbf{M}_1$ and $\mathbf{M}_2$ is that $\mathbf{M}_2 \mathbf{M}_1$$\mathbf{M}_2 \mathbf{M}_1 \neq \mathbf{I}$. Hence encircling the loop in Fig. \ref{FigExample1}(b) twice still lead to nontrivial exchange. For example, the state $s_1$ will evolve into $s_3$, $s_4$ and $s_2$ after encircling the loop two, three and four times, respectively.

\begin{figure} 
	\includegraphics[width=\linewidth]{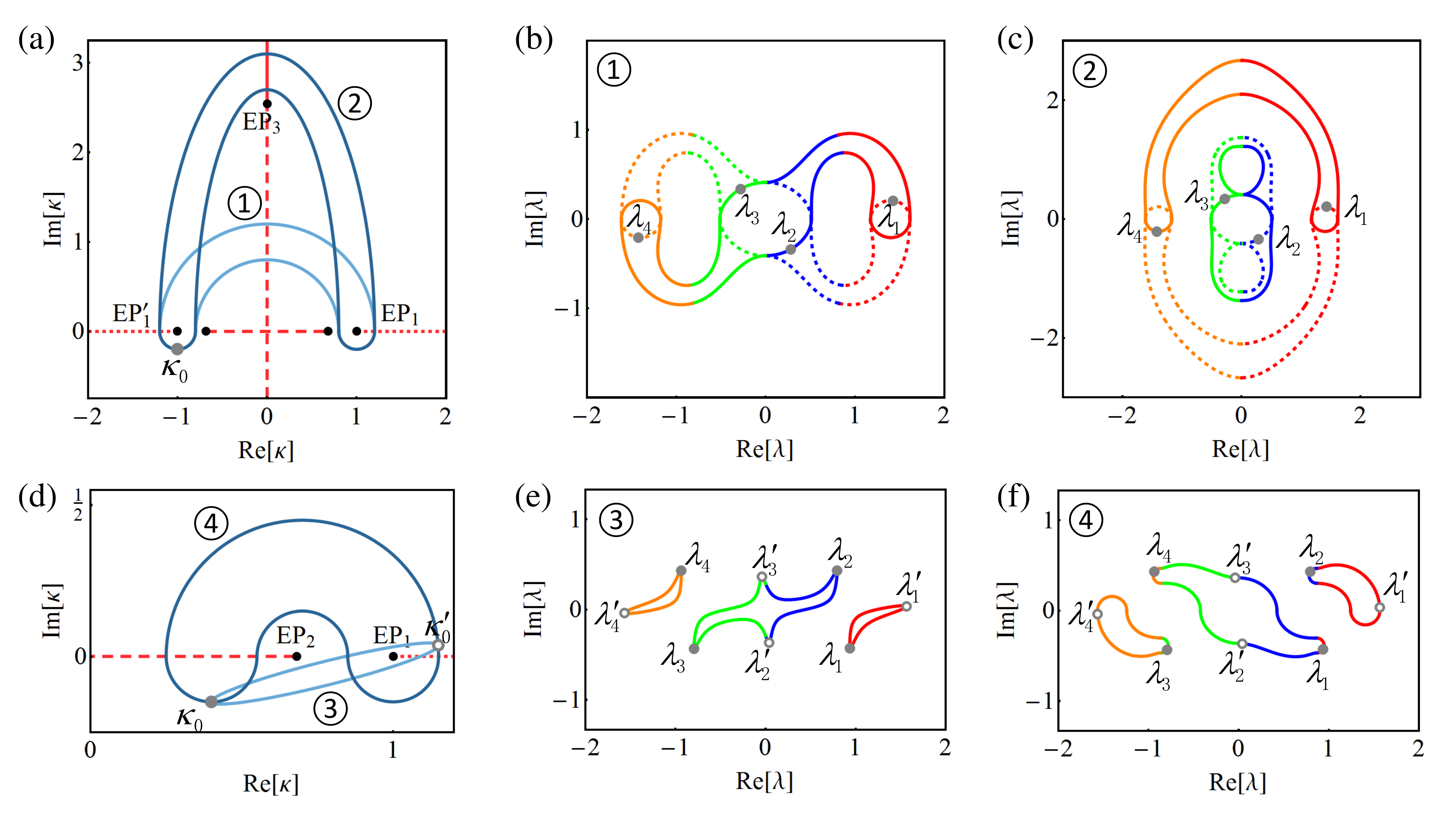}
	\caption{\textbf{Numerical example of homotopic relations between loops.} (a) Depicts two similar loops \textcircled{1} and \textcircled{2} that encircle EP$_1$ and EP$_2$. The two loops are non-homotopic for any starting point including $\kappa_0$ (which is considered for the example), since they cannot be deformed into one another without crossing $\text{EP}_3$. Their corresponding matrix product is  $\mathbf{M}_1\mathbf{M}_2\mathbf{M}_1\mathbf{M}_2$ and $\mathbf{I}$, respectively. This is confirmed by their eigenvalue trajectories as shown in (b) and (c). (d) The two similar loops  \textcircled{3} and \textcircled{4} are non-homotopic for the starting point $\kappa_0$ but homotopic for $\kappa_0'$. This is also reflected in the exchange relations of the eigenvalues as shown in (e) and (f).}
	\label{FigExample2} 
\end{figure}

\noindent \textit{Topological features of equivalent loops---} Here, we further elucidate on the topological features of equivalent loops in the context of the example given by Eq. (\ref{H}). In this case, the space $\bar{S}$ would be the space spanned by Re$[\kappa]$ and Im$[\kappa]$ after removing the points EP$_{1,2,3}$ and EP$_{1,2,3}'$. By inspecting the two loops \textcircled{1} and \textcircled{2} in Fig. \ref{FigExample2}(a), it is clear that they are not homotopic for the starting point $\kappa_0$. Indeed the net permutation matrix associated with loop \textcircled{1} is $\mathbf{M}_1\mathbf{M}_2 \mathbf{M}_1 \mathbf{M}_2$, resulting in $\{s_1, s_2,s_3, s_4\}\rightarrow \{s_4, s_3,s_2, s_1\}$. However, the permutation matrix associated with loop \textcircled{2} is $\mathbf{M}_1 \mathbf{M}_3 \mathbf{M}_1 \mathbf{M}_3=\mathbf{I}$. Consequently their exchange relations are in general different as shown in Fig. \ref{FigExample2}(b) and (c). 

Next, we investigate a scenario that highlight the case of free homotopy. The two loops \textcircled{3} and \textcircled{4} in Fig. \ref{FigExample2}(d) are similar (enclose the same EPs), yet they are not homotopic for the starting point $\kappa_0$, i.e. they cannot be transformed into one another while keeping the starting point fixed and without crossing EP$_2$. Thus the two loops are not necessarily equivalent. Indeed the net redistribution matrix associated with loop \textcircled{3} is $\mathbf{M}_1$, resulting in $\{s_1, s_2,s_3, s_4\}\rightarrow \{s_1, s_3,s_2, s_4\}$; while for loop \textcircled{4}, the permutation matrix is $\mathbf{M}_2 \mathbf{M}_1 \mathbf{M}_2$, which gives $\{s_1, s_2,s_3, s_4 \} \rightarrow \{s_4, s_2,s_3, s_1 \}$. On the other hand, if we consider the same loops \textcircled{3} and \textcircled{4} but with a different starting point $\kappa_0'$, they are homotopic and the net permutation matrix is $\mathbf{M}_1$ for both loops. Figures \ref{FigExample2}(e) and (f)  confirm these results.

\noindent \textit{Implications for dynamical evolution---}
So far we have discussed the stroboscopic (or static) exchange between the eigenstates as a result of encircling EPs. Whereas this type of ``evolution" can be in general accessed experimentally (see Refs. \cite{Dembowski2001PRL,Dietz2011PRL,Gao2015N} for the case of second order EPs), recent theoretical and experimental efforts are painting a different picture for the dynamic evolution, showing that the interplay between gain and loss will inevitably break adiabaticity \cite{Raam2011JPA,Berry2011JPA,Berry2011JO,Hassan2017PRL,Doppler2016N,Xu2016N}. It will be thus interesting to investigate whether the homotopy between the loops (or its lack for that matter) has any impact on the dynamic evolution. Here we do not attempt to answer this question rigorously but will rather consider illustrative example. To do so, we focus again on the same loops \textcircled{3} and \textcircled{4}  shown in Fig. \ref{FigExample2}(d), and we perform numerical integration to compute the dynamical evolution around these loops starting from either $\kappa_0$ or $\kappa_0'$. As we discussed before, the loops are similar for both initial conditions but homotopic only for the later one. The computational details are presented below but the main results confirm our conclusion in the main text: (1) When the two loops are homotopic (i.e when the initial point on the the loop is $\kappa_0'$) any initial state $s_i$, with $i=1,2,3,4$, will end up at state $s_2$ regardless of the considered loop; (2) For similar but non-homotopic loops (i.e when the initial point on the the loop is $\kappa_0$), the initial states on loop \textcircled{3} always evolve to $s_3$ while those on loop \textcircled{4} will evolve to $s_1$.
These results suggest that homotopy between the loops plays a much greater role than just describing the static exchange between the states. Particularly, it might be also useful in classifying the dynamic evolution. We plan to investigate this interesting direction in future work.  

\noindent\textit{Numerical calculation of dynamic evolution---} Here we present the details of the numerical calculations for the dynamic evolution. First, we choose the point $\kappa_0 = (0.4,-0.15)$ in Fig. 6. Next, choose the loop \textcircled{4} in Fig. 6(a) as:
\begin{equation}
\begin{aligned}
\text{Re}[\kappa(t)]&=\begin{cases}
c_1+r_1 \cos(\omega t), & t \in [0,T/4) \\
c_2+r_2 \cos(\omega t), & t \in [T/4,T/2) \\
c_1-r_2 \cos(\omega t), & t \in [T/2,3T/4) \\
c_3+r_2 \cos(\omega t), & t \in [3T/4,T] 
\end{cases}, \\
\text{Im}[\kappa(t)]&=\begin{cases}
r_1 \sin(\omega t), & t \in [0,T/4) \\
r_2 \sin(\omega t), & t \in [T/4,T] 
\end{cases} ,
\end{aligned}
\end{equation} 
%}
where $c_1=0.7$, $c_2=0.4$ and $c_3=1$. Note that the centers of the semicircles associated with loop \textcircled{4} in Fig. 6(a) are given by the coordinates $(c_{1,2,3},0)$. The associated radii are $r_1=0.45$ and  $r_2=0.15$. The quantity $T=4 \pi/|\omega|$ is the time needed to complete one cycle. 
The exact position of point  $\kappa_0'$ can be now chosen to be  the intersection between the line passing  through $\kappa_0$ and EP$_1$ and the top large semi-circle, and $\kappa'_0 \approx (1.148,0.03711)$. 

Finally, loop \textcircled{3} in Fig. 6(b) was chosen to be a titled ellipse  with the line connecting $\kappa_0$ and $\kappa'_0$ as the major axis. This ellipse has semi-major axis $a \approx 0.3782$, focal distance $c=a-0.002$ and a rotating angle $\theta=\arctan\frac{1}{4}$. Therefore the parametric function of loop \textcircled{3} is:
\begin{equation}
\begin{aligned}
\text{Re}[\kappa(t)]&= c_x + a \cos(\omega t) \cos\theta - b \sin(\omega t) \sin\theta,\\
\text{Im}[\kappa(t)]&= c_y + a \cos(\omega t) \sin\theta + b \sin(\omega t) \cos\theta,
\end{aligned}
\end{equation} 
where $b=\sqrt{a^2-c^2}$ is the semi-minor axis of the ellipse and $(c_x,c_y)=(c_2+a \sin\theta, -r_2 + a \cos \theta)$ is the center of the ellipse.

In all simulations, we chose the encircling speed $\omega=\pm 10^{-4}$ (the positive/negative signs CCW/CW respectively).

\begin{figure} 
	\center
	\includegraphics[width=\linewidth*3/4]{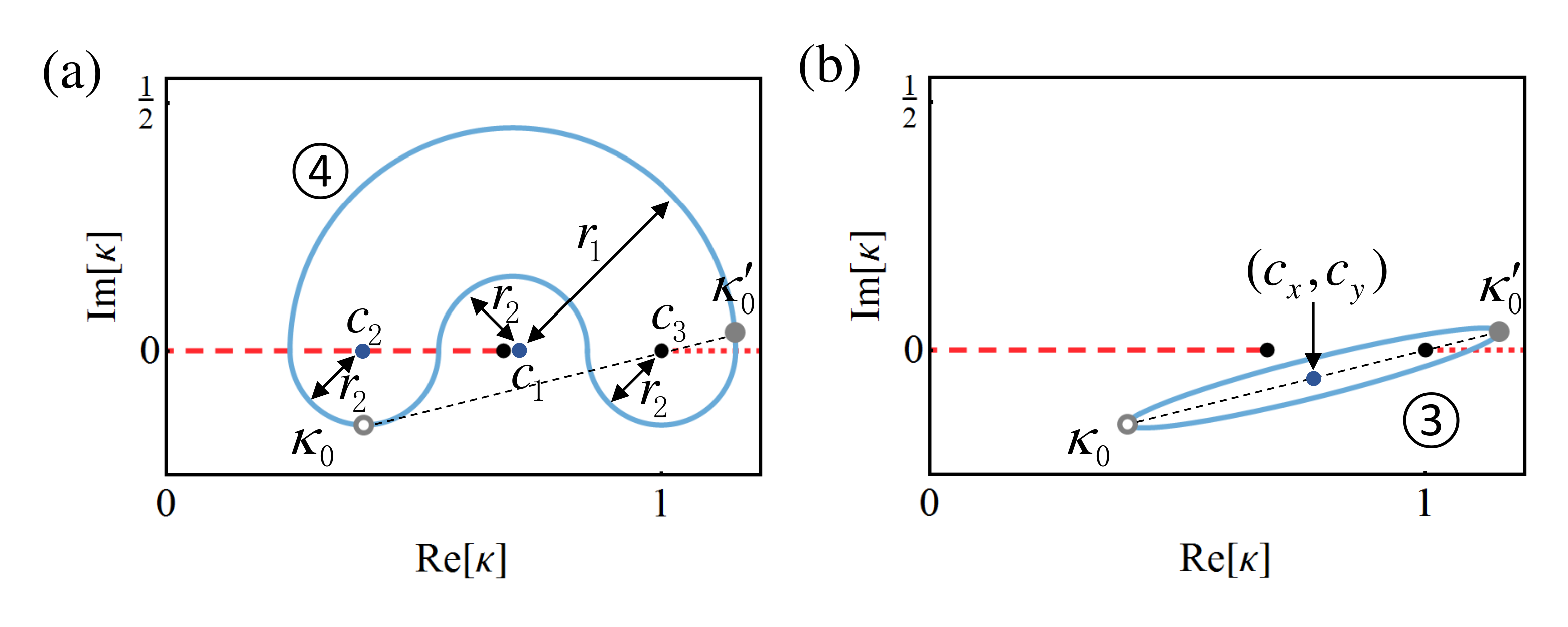}
	\caption{\textbf{Trajectories of dynamical evolutions.} The details of loops \textcircled{3} and \textcircled{4} used in the numerical simulation of dynamic evolution of eigenstates in the main text are illustrated in (a) and (b). Loop \textcircled{3} is a titled ellipse  with the line connecting $\kappa_0$ and $\kappa'_0$ as the major axis. Loop \textcircled{4} is a combination of one large semi-circle and three identical small semi-circles.}
	\label{FigNumerical} 
\end{figure}

\newpage
\noindent \textit{Data availability---} The data that support the findings of this study are available from the corresponding author upon reasonable request.

\subsection*{Author Contribution}
R.E. conceived the project. Q.Z. and R.E. developed the theoretical framework with support from D.N.C. All Authors contributed to the analysis and manuscript writing.

\bibliographystyle{naturemag.bst}
\bibliography{Reference}

\end{document}